\documentclass[reprint,amsmath,amssymb,aps,prl,longbibliography,superscriptaddress,noeprint,nofootinbib]{revtex4-1}

\usepackage[pdftex]{graphicx}
\usepackage{wasysym}
\usepackage{ifsym}
\usepackage{collref}
\usepackage[left=1.9cm,right=1.9cm,top=2.9cm,bottom=2.9cm]{geometry}
\usepackage[utf8]{inputenc}
\usepackage[english]{babel}
\usepackage{amsmath}
\usepackage{verbatim}   
\usepackage{subfigure}  
\usepackage{amsfonts}
\usepackage{amssymb}
\usepackage{mathrsfs}
\usepackage{stmaryrd}
\usepackage{slashed}
\usepackage{amsthm}
\usepackage{dcolumn}
\usepackage{bm}
\usepackage[usenames,dvipsnames]{xcolor}
\usepackage{extarrows}   
\usepackage{cleveref}
\usepackage[]{units} 
\usepackage{tikz}
\usepackage{scalerel}
\usepackage{simplewick}
\usepackage{mathtools}
\usepackage{float}

\def\ch{\text{ch}} 
\def\td{\text{td}}
\def\SL{\text{SL}}
\def\Ell{\text{Ell}}

\newcommand\myeq{\stackrel{\mathclap{\normalfont\mbox{\scriptsize def}}}{=}}

\begin{document}

\vspace*{-2cm}
\begin{flushright}
	{KIAS-P18089}
\end{flushright}

\title{Algebraic surfaces, Four-folds and Moonshine}

	\author{Kimyeong Lee}
	\email{klee@kias.re.kr}
	\affiliation{School of Physics, Korea Institute for Advanced Study, Seoul 02455, Korea}
	
	\author{Matthieu Sarkis}
	\email{sarkis@kias.re.kr}
	\affiliation{School of Physics, Korea Institute for Advanced Study, Seoul 02455, Korea}

\begin{abstract}

	The aim of this note is to point out an interesting fact related to the elliptic genus of complex algebraic surfaces in the context of Mathieu moonshine. We also discuss the case of 4-folds.

\end{abstract}

\maketitle

\section{Introduction}

	In their seminal paper \cite{Eguchi:2010ej}, Eguchi, Ooguri and Tachikawa made the interesting observation that the elliptic genus of K3 admits a decomposition in terms of $\mathcal N=4$ super-Virasoro characters, whose coefficients happen to correspond to the dimension of representations of the largest Mathieu sporadic group $\mathbb M_{24}$. This phenomenon was formalized in a precise conjecture related to the existence of a graded module for $\mathbb M_{24}$ whose properties would precisely mimic the experimental observation of \cite{Eguchi:2010ej}. This conjecture was then proved by Gannon \cite{Gannon:2012ck}. Many authors have made important contributions towards understanding Mathieu moonshine from a physics perspective in terms of the BPS spectrum of $\mathcal N=(4,4)$ NSLM with K3 target. 
	
	In parallel to that, Eguchi and Hikami \cite{Eguchi:2013es} made the interesting observation that since K3 surfaces correspond to a double cover of Enriques surfaces, the elliptic genus of the latter is given by
	\begin{equation}
		\Ell_{\text{Enriques}}=\frac{1}{2}\,\Ell_{\text{K3}}=\phi_{0,1}\,,
	\end{equation}
    where $\phi_{0,1}$ is the standard weight-0 index-1 weak Jacobi form (\ref{eq:def_stdJacobiForm}). They then noticed that the evenness of the coefficients in the $\mathcal N=4$ expansion of $\Ell_{\text{K3}}$ allowed them to enlighten a connection between $\Ell_{\text{Enriques}}=\phi_{0,1}$ and the Mathieu group $\mathbb M_{12}$. They however underlined the fact that since the Enriques surfaces are not Calabi-Yau, they cannot define naively a genuine string background, hence making more obscure any string theoretic explanation of the observed $\mathcal N=4$ decomposition. This result, however, seems to indicate that from the geometric perspective the moonshine phenomena are not restricted to complex surfaces admitting a Ricci-flat Riemannian metric. 
	
	In this note, we consider the case of a generic smooth compact complex algebraic surface $X$. Even in absence of a justifiable field theoretic definition and derivation of the elliptic genus (by localization using a GLSM realization when $X$ is toric for instance), one can consider the mathematical definition of the elliptic genus given in terms of the holomorphic Euler characteristic of a formal series with holomorphic vector bundle coefficients. We give for it a simple expression in terms the self-intersection number of the canonical class and the Euler number. One can then deduce the behaviour of the elliptic genus under blow ups, allowing us to state in which sense any surface with positive $K^2$ may be relevant for a geometric understanding of Mathieu moonshine.
	
	We finally discuss how the discussion for surfaces can be extended for 4-folds.\newline

	\textsc{Conventions:}
		\begin{itemize}
			\item We will omit the modular argument and denote by $\theta(z)$ the odd Jacobi theta function $\theta(\tau,z)$ when unambiguous, cf. appendix.
			\item $q=\exp(2i\pi\tau)$ and $y=\exp(2i\pi z)$. Differentiation of the odd Jacobi theta function is always with respect to the elliptic variable $z$.
		\end{itemize}
					
\section{Elliptic genus of complex algebraic surfaces}
\label{sec:GeomDerivation}
	
	In the following we denote by $X$ a smooth compact complex algebraic surface. We will also formally write
	\begin{equation}
		c(X)=\left(1+x_1\right)\left(1+x_2\right)
	\end{equation}
	for the total Chern class. We define the following formal series with bundle coefficients:
	\begin{equation}
	\label{eq:formalbundle}
		\scaleto{\mathbb{E}_{q,y} = y^{-1}\bigotimes_{n=1}^\infty\left(\Lambda_{-y q^{n-1}} T^\star_X \otimes\Lambda_{-y^{-1} q^n} T_X \otimes S_{q^n} T^\star_X \otimes S_{q^n} T_X\right)\, ,}{25.5pt}
	\end{equation}
	where we defined the total exterior and symmetric powers
	\begin{equation}
		\Lambda_{t} E = \sum_{k=0}^{\infty}\, t^k\, \Lambda^k E\ , \quad S_t E =\sum_{k=0}^{\infty}\, t^k \, S^k \, E\, ,
	\end{equation}
	$\Lambda^k$ and $S^k$ being respectively  the $k$-th exterior product and the $k$-th symmetric product. The elliptic genus is then given by the holomorphic Euler characteristic of this holomorphic vector bundle $\mathbb{E}_{q,y}$, which, using Hirzebruch-Riemann-Roch theorem, is equal to:
	\begin{equation}
		\Ell_X(\tau,z)=\int_X\ch(\mathbb{E}_{q,y})\,\td(X)\,.
	\end{equation}
	The total Chern character is well-behaved with respect to exterior and symmetric powers:
	\begin{subequations}
		\begin{align}
			\ch(S_t T_X)&=\prod_a\frac{1}{1-te^{x_a}}\,,\\ \ch\left(\Lambda_{t} T_X\right)&=\prod_a\left(1+te^{x_a}\right)\,.
		\end{align}
	\end{subequations}
	Using \cref{eq:def_theta}, we then obtain:
	\begin{equation}
		\Ell_X(\tau,z)=\int_X\prod_ax_a\frac{\theta\left(\frac{x_a}{2i\pi}-z\right)}{\theta\left(\frac{x_a}{2i\pi}\right)}\,.
	\end{equation}
	Moreover we can Taylor expand as follows
	\begin{equation}
		\theta(z+u)=\theta(z)\exp\left(\sum_{n=1}^\infty\frac{u^n}{n!}\frac{\partial^n}{\partial z^n}\log\theta(z)\right)
	\end{equation}
	We are therefore left with
	\begin{equation}
		\begin{split}
			&\Ell_X(\tau,z)=\theta(z)^2\int_X\frac{x_1}{\theta\left(\frac{x_1}{2i\pi}\right)}\frac{x_2}{\theta\left(\frac{x_2}{2i\pi}\right)}\times\\
			&\times\exp\left(\sum_{n=1}^\infty\frac{1}{n!}\frac{\partial^n\log\theta(z)}{\partial z^n}\sum_a\left(-\frac{x_a}{2i\pi}\right)^n\right)\,.
		\end{split}	
	\end{equation}	
	One can show that 
	\begin{equation}
	\label{eq:identity_Eisenstein}
		\frac{\theta^{(3)}(0)}{\theta'(0)}=-3E_2(\tau)\,,
	\end{equation}		
	where $E_2$ is the weight-2 $\text{SL}_2(\mathbb Z)$ Eisenstein series normalized as
	\begin{equation}
		E_2(\tau)=\sum_{\omega\in\mathbb Z+\tau\mathbb Z}'\frac{1}{\omega^2}\,,
	\end{equation}
	the prime on the sum meaning that the origin of the lattice is omitted in the sum.
		
	Using \cref{eq:identity_Eisenstein}, collecting the contribution of the non exponentiated piece, truncating the Taylor series up to order 2, which is possible since we work under the integral, and taking care of the $2i\pi$'s, we obtain
	\begin{equation}
		\begin{split}
			\Ell_X(\tau,z)&=\left(\frac{2i\pi\theta(z)}{\theta'(0)}\right)^2\int_X\exp\left\{-\frac{\partial\log\theta(z)}{\partial z}\sum_a\frac{x_a}{2i\pi}+\right.\\
			&\left.+\frac{1}{2}\left(\frac{\partial^2\log\theta(z)}{\partial z^2}+E_2(\tau)\right)\sum_a\left(\frac{x_a}{2i\pi}\right)^2\right\}\,.
		\end{split}
	\end{equation}
	Expanding to top form order, we get
	\begin{equation}
		\begin{split}
			\Ell_X(\tau,z)&=\left(\frac{\theta(z)}{\theta'(0)}\right)^2\int_X\exp\left\{-\frac{\partial\log\theta(z)}{\partial z}\,\ch_1(X)+\right.\\
			&\left.+\left(\frac{\partial^2\log\theta(z)}{\partial z^2}+E_2(\tau)\right)\,\ch_2(X)\right\}\,.
		\end{split}
	\end{equation}	
	One can easily show that the following identity holds:
	\begin{equation}
		\frac{\partial\log\theta(z)}{\partial z}=\zeta_w(\tau,z)-E_2(\tau)z\,,
	\end{equation}
	where $\zeta_w$ is the Weierstrass elliptic zeta function, satisfying $\partial_z\zeta_w(\tau,z)=-\wp(\tau,z)$. 
	
	\begin{comment}
	Indeed, the odd Jacobi theta function has the infinite product representation:
	\begin{equation}
		\theta(\tau,z)=2q^\frac{1}{8}\sin(z)\prod_{n=1}^\infty(1-q^n)(1-yq^n)(1-y^{-1}q^n)\,,
	\end{equation}
	from which we get
	\begin{equation}
		\frac{\partial\log\theta(z)}{\partial z}=\cot(z)-2i\pi\sum_{n=1}^\infty\frac{(y-y^{-1})q^n}{(1-yq^n)(1-y^{-1}q^n)}\,,
	\end{equation}
	which we can rewrite
	\begin{equation}
		\frac{\partial\log\theta(z)}{\partial z}=-i\pi\sum_{n\in\mathbb Z}\frac{1-y^2}{(1-yq^n)(1-y^{-1}q^n)}\,,
	\end{equation}
	from which we obtain the required result
	\begin{equation}
		\frac{\partial\log\theta(z)}{\partial z}&=\zeta_w(\tau,z)-E_2(\tau)z\,.
	\end{equation}
	\end{comment}
	
	We therefore end up with the following expression for the elliptic genus in terms of the self-intersection number $K^2$ of the canonical class and the Euler number $e$:
	\begin{equation}
	\label{eq:GenericComplexSurface}
		\begin{split}
			&\Ell_X(\tau,z)=\left(\frac{\theta(z)}{\theta'(0)}\right)^2\times\\
			&\times\left\{-\left(\frac{K^2}{2}-e\right)\,\wp(\tau,z)+\frac{K^2}{2}\,\left(\zeta_w(\tau,z)-E_2(\tau)z\right)^2\right\}\,.
		\end{split}
	\end{equation}
	
	Using the multiplicative property of the elliptic genus, one can interprete $-2\left(\zeta_w(\tau,z)-E_2(\tau)z\right)$ as the elliptic genus of $\mathbb P^1$.
	
	Before proceeding, let us mention as a parenthesis that, as one could expect, since differentiation with respect to the elliptic variable is not covariant in the sense of Jacobi forms, $\Ell_X$ is generically not a Jacobi form. Drawing lessons from \cite{2015arXiv150403787R}, we introduce the canonical raising operator, acting on the odd Jacobi theta function as:
	\begin{equation}
		Y_+^{\frac{1}{2},\frac{1}{2}}(\theta)(\tau,z)=\theta'(\tau,z)+2i\pi\,\frac{z-\bar z}{\tau-\bar\tau}\,\theta(\tau,z)\,,
	\end{equation}
	which is then a genuine Jacobi form, but at the cost of holomorphicity both in the modular and elliptic arguments. We are therefore tempted to replace the logarithmic derivative $\partial\log\theta/\partial z$ in (\ref{eq:GenericComplexSurface}) by its covariant avatar. One can therefore canonically associate to any compact algebraic surface the following weight-0 index-1 non-holomorphic elliptic genus:
	\begin{equation}
	\label{eq:ModComplGenus}
		\begin{split}
		&\widetilde\Ell_X(\tau,z)\myeq\left(\frac{\theta(z)}{\theta'(0)}\right)^2\left\{-\left(\frac{K^2}{2}-e\right)\,\wp(\tau,z)+\right.\\
		&\left.+\frac{K^2}{2}\,\left(\zeta_w(\tau,z)-E_2(\tau)z+2i\pi\,\frac{z-\bar z}{\tau-\bar\tau}\right)^2\right\}\,.
		\end{split}
	\end{equation}
	Considering the elliptic genus as a building block capturing part of the physics of branes wrapping the surface $X$ in, say, an M or F-theory framework, it would be very interesting to understand the physical meaning of such an non-holomorphy in this compact setting\footnote{We thank A. Libgober for pointing out earlier work describing this non-holomorphicity in terms of the quasi-Jacobi nature of the elliptic genus \cite{2009arXiv0904.1026L}.}.	
		
\section{Moonshine phenomena}
		
	\subsection{Behaviour of the elliptic genus of surfaces under blow-ups}
	
		As a very nice corollary of the generic formula (\ref{eq:GenericComplexSurface}) we can learn the behaviour of the elliptic genus under blow-ups. Indeed, denoting by $\tilde X$ the blown-up surface at an arbitrary point, $\pi$ the blow-up morphism, $E$ the exceptional divisor, and $D,D'$ two divisors on $X$, we have the following classical results for the intersection pairing: 
		\begin{align}
			&E\cdot E=-1\,,\nonumber\\ 
			&\pi^*D\cdot\pi^*D'=D\cdot D'\,,\\ 
			&\pi^*D\cdot E=0\nonumber\,,
		\end{align}
		from which we obtain that 
		\begin{equation}
			K_{\tilde X}^2=K_X^2-1\,.
		\end{equation}
		Since the holomorphic Euler characteristic of the structure sheaf is a birational invariant, Noether's formula then tells us that $K^2+e$ is also a birational invariant of surfaces, hence $e_{\tilde X}=e_X+1$.
		
		We can finally deduce the behaviour of the elliptic genus under the blow-up at an arbitrary point:
		\begin{equation}
			\begin{split}
				\Ell_{\tilde X}(\tau,z)=&\ \Ell_X(\tau,z)+\left(\frac{\theta(z)}{\theta'(0)}\right)^2\times\\
				&\times\left\{-\frac{1}{2}\left(\zeta_w(\tau,z)-E_2(\tau)z\right)^2+\frac{3}{2}\,\wp(\tau,z)\right\}\,.
			\end{split}
		\end{equation}
		The correction term can be interpreted as the elliptic genus of a surface with $K^2=-1$ and $e=1$:
		\begin{equation}
			\Ell_{\tilde X}=\Ell_X+\Ell_{(K^2=-1\,;\,e=1)}\,,
		\end{equation}	
		corresponding to the fact that in the blow-up process on removes a point (with $e(\text{pt})=1$) to replace it by the $\mathbb P^1$ of all outward directions through the point (with $e\left(\mathbb P^1\right)=2$).	
		 We therefore find the following very interesting result:
		\newline
		
       \textit{In addition to $K3$ and Enriques surfaces, minimal surfaces of Kodaira dimension 1, namely proper elliptic surfaces, have an elliptic genus proportional to $\phi_{0,1}$.} 
       \newline
       
       \textit{Moreover, one can start from any complex algebraic surface with topological numbers $(K^2,e)$ such that $K^2\geq0$, blow it up at $K^2$ generic points to get a surface with Euler number $e+K^2$, whose elliptic genus will also be proportional to $\phi_{1,0}$, hence is also related  to at least the Mathieu group $\mathbb M_{12}$.}
       \newline
       
       As a simple remark, let us add that surfaces with Euler number congruent to $0$ mod $24$ will further exhibit $\mathbb M_{24}$ moonshine.
       
       Let us now illustrate this generic statement by some examples.
       	
	\subsection{One simple example: projective plane blown-up at nine points}	

    	Starting from the simplest surface and following the principle described above, we blow-up the projective plane at nine arbitrary points. The elliptic genus of the obtained surface is 
    	\begin{equation}
    		\Ell_{\text{Bl}_9(\mathbb P^2)}=\phi_{0,1}=\Ell_{\text{Enriques}}
    	\end{equation}
    	
    	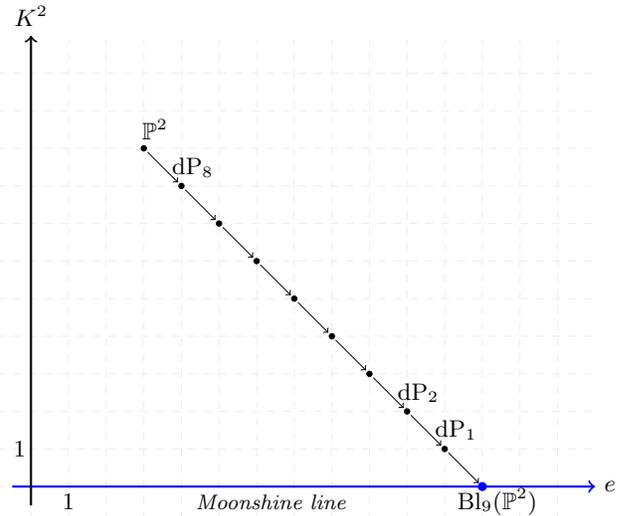
\begin{figure}[H]
    	\label{fig:Sequence_blow-ups}
        	\centering{
            	\begin{tikzpicture}[scale=0.50]
            		\draw[help lines, color=gray!15, dashed] (-0.9,-0.9) grid (14.9,11.9);
            		\draw[->,thick,blue] (-0.5,0)--(15,0) node[right,black]{$e$};
            		\draw[->,thick] (0,-0.5)--(0,12) node[above]{$K^2$};
    		
            		\filldraw [black] (3,9) circle (2pt);
    				\filldraw [black] (4,8) circle (2pt);
    				\filldraw [black] (5,7) circle (2pt);
    				\filldraw [black] (6,6) circle (2pt);
    				\filldraw [black] (7,5) circle (2pt);
    				\filldraw [black] (8,4) circle (2pt);
    				\filldraw [black] (9,3) circle (2pt);
    				\filldraw [black] (10,2) circle (2pt);
    				\filldraw [black] (11,1) circle (2pt);
    				\filldraw [blue] (12,0) circle (3pt);
    				
    				\draw[->,ultra thin] (3.1,8.9)--(3.9,8.1);
    				\draw[->,ultra thin] (4.1,7.9)--(4.9,7.1);
    				\draw[->,ultra thin] (5.1,6.9)--(5.9,6.1);
    				\draw[->,ultra thin] (6.1,5.9)--(6.9,5.1);
    				\draw[->,ultra thin] (7.1,4.9)--(7.9,4.1);
    				\draw[->,ultra thin] (8.1,3.9)--(8.9,3.1);
    				\draw[->,ultra thin] (9.1,2.9)--(9.9,2.1);
    				\draw[->,ultra thin] (10.1,1.9)--(10.9,1.1);
    				\draw[->,ultra thin] (11.1,0.9)--(11.9,0.1);		
    				
					\node at (1,-0.4) {$1$};
					\node at (-0.3,1) {$1$};
    				\node at (3.3,9.5) {$\mathbb P^2$};
    				\node at (4.3,8.5) {$\text{dP}_8$};
    				\node at (10.3,2.5) {$\text{dP}_2$};
    				\node at (11.3,1.5) {$\text{dP}_1$};
    				\node at (12.4,-0.4) {$\text{Bl}_9(\mathbb P^2)$};
					\node at (6.4,-0.4) {\footnotesize{\textit{Moonshine line}}};
        		\end{tikzpicture}
    		}
    	\caption{Blow-up orbit of $\mathbb P^2$ towards the moonshine line.}
    	\end{figure}
    	
		The projective plane blown-up in nine points is also called $\frac{1}{2}$K3 and appeared in \cite{Cheng:2017dlj} as the base of an elliptically fibered Calabi-Yau 3-fold geometry defining a genuine supersymmetric string background. Configurations with fivebranes wrapping the base inside the CY$_3$ were studied there, and a connexion was made with $\mathbb M_{12}$. This is in nice accordance with the claim of our note, which gives a more basic reason of why the blow-up of $\mathbb P^2$ in nine points should be related the small Mathieu group $\mathbb M_{12}$. Notice that in the $(K^2,e)$ plane, Enriques surfaces sit on the same point as $\text{Bl}_9(\mathbb P^2)$.
	
		It would be extremely interesting if the geometry of the blown-up projective plane could give us insights concerning the moonshine phenomenon.
		
    	In particular, the automorphism group of the graph of exceptional curves of the blow-up of $\mathbb P^2$ in nine points in general position is discrete and is related to the Weyl group of the affine extension $E_9$ of the $E_8$ Dynkin diagram. A classical result on affine Weyl groups then tells us that the latter is given by the semi-direct product of the Weyl reflection group of the $E_8$ Dynkin diagram and translations in the co-root lattice $W(E_9)=W(E_8)\rtimes\mathbb Z^8$. In addition, it is known that the Weyl group of the $E_8$ Dynkin diagram is given in terms of the classical Chevalley group $\Omega_8^+(2)$. One may want to try and relate a 'sigma-model theoretic' extension of that automorphism group to $\mathbb M_{12}$. This will be the content of a future work.	
	
	\subsection{Other examples}
	
		Let us simply mention a few other examples. In the same spirit as the projective plane, one may also consider starting from any of the Hirzebruch surface $\mathbb F_n$, and blow it up in $8$ points. The obtained surface will then exhibit $\mathbb M_{12}$ moonshine.
		
		For higher Euler characteristic, it was shown by Hirzebruch and Van de Ven \cite{HirzebruchHilbert} that the Hilbert modular surfaces $Y^0(p)$ for $p\in\{53,61,73\}$ are proper elliptic surfaces. These three surfaces satisfy $K_{Y^0(p)}^{\,2}=0$ and have Euler number $e_{Y^0(p)}=36$. Their elliptic genus is therefore given by $\Ell_{Y^0(p)}=3\,\phi_{0,1}$, hence exhibit $\mathbb M_{12}$ moonshine.  
		
		Finally, let us give as a last example a surface with Euler number equal to $48$, hence connected to $\mathbb M_{24}$. Consider a double covering of the projective plane, ramified over a curve of degree $8$ with at most simple singularities, perform the canonical resolution of the possible singularities. This resolved double covering $\bar{\mathbb P}^2$ of the projective plane is a minimal surface of general type. Finally blow-up the obtained smooth surface at two arbitrary points. We claim that the resulting surface has $\Ell_{X}=4\,\phi_{0,1}$, as wanted. 
		
		We therefore see that considering coverings allows to build new surfaces relevant for Mathieu moonshine. K3 surfaces being double coverings of Enriques surfaces appear therefore here as a particular case of more generic constructions involving coverings blown-up at various points. 
		
		\begin{figure}[H]
    	\label{fig:a}
        	\centering{
            	\begin{tikzpicture}[scale=0.425]
            		\draw[help lines, color=gray!15, dashed] (-0.9,-0.9) grid (17.9,3.9);
            		\draw[->,thick] (-0.5,0)--(18,0) node[right]{$\frac{e}{12}$};
            		\draw[->,thick] (0,-0.5)--(0,4) node[above]{$K^2$};
    		
   					\filldraw [black] (15.3,2) circle (2pt);
    				\filldraw [blue] (4,0) circle (3pt);
				
					\filldraw [blue] (8,0) circle (3pt);
					\filldraw [blue] (12,0) circle (3pt);
					\filldraw [blue] (16,0) circle (3pt);
    				
					\draw[->,ultra thin, dashed] (15.4,1.8)--(15.9,0.2);
				    				
					\node at (1,-0.5) {$\frac{1}{4}$};
					\node at (-0.3,1) {$1$};
					\node at (15.3,2.5) {$\bar{\mathbb P}^2$};
    				\node at (4.4,-0.5) {$\text{Bl}_9(\mathbb P^2)$};
					\node at (4.4,-1.5) {$\text{Bl}_8(\mathbb F_n)$};
					\node at (4.4,-2.5) {Enriques};
					\node at (8,-0.5) {K3};
					\node at (12.4,-0.5) {$Y^0(p)$};
					\node at (16.2,-0.5) {$\text{Bl}_2\left(\bar{\mathbb P}^2\right)$};
        		\end{tikzpicture}
    		}
    	\caption{Examples of moonshine surfaces at $e=12,24,36,48$.}
    	\end{figure}
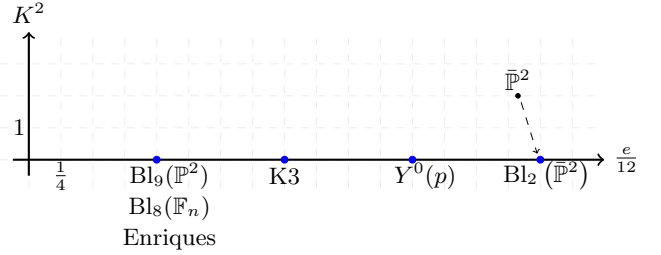
		
\section{4-folds}
	
	Let $X_4$ be a compact algebraic 4-fold. Upon use of (\ref{eq:various_identities}), a longer computation gives us:
	\begin{equation}
	\label{eq:EllGen4fold}
		\begin{split}
			&\Ell_{X_4}(\tau,z)=\left\{-6\left(\wp^2-6E_4\right)\ch_4+\frac{1}{2}\wp^2\,\ch_2^2-\right.\\
			&-\left(\zeta_w-E_2z\right)\wp'\,\ch_1\ch_3-\frac{1}{2}\left(\zeta_w-E_2z\right)^2\wp\,\ch_1^2\ch_2+\\
			&\left.+\frac{1}{24}\left(\zeta_w-E_2z\right)^4\ch_1^4\right\}\left(\frac{\theta(z)}{\theta'(0)}\right)^4\,.
		\end{split}
	\end{equation}
	
	In \cite{Eguchi:2012ye}, Eguchi and Hikami uncovered an $\mathcal N=2$ moonshine phenomenon for $\text{L}_2(11)$, the projective group of special linear matrices with coefficients in a finite field with 11 elements, this time in relation with some specific Calabi-Yau four-folds.
	
	We conclude from this:\newline
	
	 \textit{All compact complex algebraic 4-folds with the following intersection numbers:}
	\begin{equation}
	\label{eq:EulerNumb}
		\begin{split}
			&\ch_1\cdot\ch_3=\ch_1^2\cdot\ch_2=\ch_1^4=0\,,\\
			&\ch_4=-10\,a\,,\ \ \ \ch_2^2=-96\,a\,,\ \ \ \text{with }a\in\mathbb Q\,,
		\end{split}
	\end{equation}
	\textit{exhibit at least the same} $\text{L}_2(11)$ \textit{moonshine phenomenon.}
	\newline
	
	Indeed, for 4-folds satisfying (\ref{eq:EulerNumb}), the elliptic genus is given by
	\begin{equation}
		\Ell_{X_4}(\tau,z)=a\,Z_1(\tau,z)\,,
	\end{equation}
	with
	\begin{equation}
		Z_1(\tau,z)=\frac{1}{12}\,\phi_{0,1}(\tau,z)^2-360\left(\frac{\theta(z)}{\theta'(0)}\right)^4E_4(\tau)\,,
	\end{equation}
	which was precisely the weight-0 index-2 Jacobi form appearing in \cite{Eguchi:2012ye}.
	
	In a similar spirit, it was shown in \cite{Cheng:2014owa} that the Jacobi form
	\begin{equation}
		Z_2(\tau,z)=\frac{1}{6}\,\phi_{0,1}(\tau,z)^2+3600\left(\frac{\theta(z)}{\theta'(0)}\right)^4E_4(\tau)\,,
	\end{equation} 
	exhibits Mathieu moonshine for $\mathbb M_{22}$, $\mathbb M_{23}$ or $\mathbb M_{24}$ when expanded in a basis of $\mathcal N=4$, $\mathcal N=2$ or extended $\mathcal N=1$ super-Virasoro characters respectively. We therefore conclude that:\newline
	
	 \textit{All compact complex algebraic 4-folds with the following intersection numbers:}
	\begin{equation}
		\begin{split}
			&\ch_1\cdot\ch_3=\ch_1^2\cdot\ch_2=\ch_1^4=0\,,\\
			&\ch_4=100\,a\,,\ \ \ \ch_2^2=1248\,a\,,\ \ \ \text{with }a\in\mathbb Q\,,
		\end{split}
	\end{equation}
	\textit{exhibit Mathieu moonshine upon expanding their elliptic genus in a super-Virasoro character basis.}
	\newline
	
	It is an interesting question to investigate whether starting from a given compact 4-fold, say $\mathbb P^4$, one can find a birationally equivalent 4-fold exhibiting $\text{L}_2(11)$ or Mathieu moonshine. A novelty with respect to the case of surfaces is the appearance, in addition to blow-ups, of new surgery operations, namely flips \cite{Borisov:2000fg}. This will be investigated in a future work.
	
\section{Conclusion}

	In this short note we tried to argue that it may be relevant to consider other algebraic surfaces in trying to understand the Mathieu moonshine phenomenon from a geometric perspective.
	
	It would be extremely interesting to embbed the other surfaces obtained by $K^2$ blow-ups, or Kodaira dimension 1 minimal surfaces in a string, M or F-theoretic framework and to study the physics of branes wrapping them. 
	
	Concerning Calabi-Yau four-folds, the discussion of our note may extend to generic compact algebraic 4-folds, for which a well chosen sequence of birational transformations could relate them to the relevant weight-0 index-2 Jacobi form exhibiting either $\text{L}_2(11)$, Mathieu or any other to be discovered moonshine. In the 2-dimensional case, the surfaces sharing the same $(K^2,e)$ as those sitting on the blow-up orbit of $\mathbb P^2$, namely del Pezzo, $\text{Bl}_9(\mathbb P^2)$ or Hirzebruch surfaces, appear to be relevant for physical constructions. One may therefore consider embedding elliptically-fibered Calabi-Yau 5-folds over the 4-folds sitting on the orbit of $\mathbb P^4$ into an F-theoretic framework, and study the corresponding effective 2d $(0,2)$ Gauged Linear Sigma Model, in the spirit of the work initiated by \cite{Schafer-Nameki:2016cfr}. This is left for later work.
	
	It will also be interesting to see whether some other finite simple groups, naturally related to $\text{L}_2(11)$ may appear in the context of 4-folds, the same way both $\mathbb M_{12}$ and $\mathbb M_{24}$ occur for surfaces.
	
	As a remark, let us finally notice that formula (\ref{eq:GenericComplexSurface}) and (\ref{eq:EllGen4fold}) can be given a meaning even for algebraic surfaces defined over an arbitrary field $k$, suggesting a definition of the elliptic genus in the framework of the algebraic cobordism theory of Levine and Morel \cite{LEVINE2001723}.
	
\subsection*{Acknowledgements}

	KL is supported in part by the National Research Foundation of Korea Grant NRF-2017R1D1A1B06034369.
	MS would like to thank Fr\'ed\'eric Campana, Hoil Kim, Thomas Goller and Shamit Kachru for useful discussions.
\appendix

\section*{Automorphic forms}
\label{app:functions}

	The Weierstrass zeta function is defined by:
	\begin{equation}
		\zeta_w(\tau,z)=\frac{1}{z}+\sum_{\omega\in\mathbb Z+\tau\mathbb Z}'\left\{\frac{1}{z-\omega}+\frac{1}{\omega}+\frac{z}{\omega^2}\right\}\,,
	\end{equation}
	the prime meaning that the lattice origin is omitted in the sum.
	The derivative of $\zeta_w$ with respect to the elliptic argument is the Weierstrass $\wp=-\partial_z\zeta_w$ function:
	\begin{equation}
		\wp(\tau,z)=\frac{1}{z^2}+\sum_{\omega\in\mathbb Z+\tau\mathbb Z}'\left\{\frac{1}{(z-\omega)^2}-\frac{1}{\omega^2}\right\}\,.
	\end{equation}
	The odd Jacobi theta function has the following infinite product representation:
	\begin{equation}
	\label{eq:def_theta}
		\theta(\tau,z)=2q^\frac{1}{8}\sin(z)\prod_{n=1}^\infty(1-q^n)(1-yq^n)(1-y^{-1}q^n)\,.
	\end{equation}
	The unique weight-0 index-1 weak Jacobi form (up to scaling) is given in terms of the Weierstrass $\wp$ function and the odd Jacobi theta function by:
	\begin{equation}
	\label{eq:def_stdJacobiForm}
		\phi_{0,1}(\tau,z)=12\left(\frac{\theta(\tau,z)}{\theta'(\tau,0)}\right)^2\wp(\tau,z)\,.
	\end{equation}
	The weight-2 and weight-4 $\SL_2(\mathbb Z)$ Eisenstein series are normalized as follows:
	\begin{equation}
		E_2(\tau)=\sum_{\omega\in\mathbb Z+\tau\mathbb Z}'\frac{1}{\omega^2}\,,\ \ \ \ \ E_4(\tau)=\sum_{\omega\in\mathbb Z+\tau\mathbb Z}'\frac{1}{\omega^4}\,.
	\end{equation}
	The weight-2 Eisenstein series is the prototypical example of a so-called Mock modular form, whose lack of modularity can be cured by the addition of non-holomorphic piece.
	
	Let us finally gather the main identities used here:
	\begin{subequations}
	\label{eq:various_identities}
		\begin{align}
			\frac{\partial\log\theta(\tau,z)}{\partial z}&=\zeta_w(\tau,z)-E_2(\tau)z\,,\\
			-\frac{1}{3}\frac{\theta^{(3)}(\tau,0)}{\theta'(\tau,0)}&=E_2(\tau)\,,\\
			\frac{1}{15}\frac{\theta^{(5)}(\tau,0)}{\theta'(\tau,0)}&=E_2(\tau)^2-2E_4(\tau)\,.
		\end{align}
	\end{subequations}
	One can also prove the following identity:
	\begin{equation}
		-\frac{1}{105}\frac{\theta^{(7)}(\tau,0)}{\theta'(\tau,0)}=E_2(\tau)^3-6E_4(\tau)E_2(\tau)+8E_6(\tau)\,,
	\end{equation}
	which would be relevant in the case of 6-folds.

\bibliography{biblioJ}

\end{document}